\documentclass[journal]{IEEEtran}

\usepackage{cite}
\usepackage{amsmath,amssymb,amsfonts}
\usepackage{textcomp}
\usepackage{graphicx}
\usepackage{tabularx}

\usepackage{newtxtext}
\usepackage{newtxmath}

\usepackage[ruled,vlined,linesnumbered]{algorithm2e}

\usepackage{float}
\usepackage{caption}
\usepackage{subcaption}

\usepackage{tikz}
\usetikzlibrary{
    shapes.geometric,
    shapes,
    arrows.meta,
    positioning,
    fit,
    calc
}

\newtheorem{theorem}{Theorem}
\newtheorem{proposition}{Proposition}
\newtheorem{lemma}{Lemma}
\newtheorem{corollary}{Corollary}
\newtheorem{remark}{Remark}

\begin{document}

\title{System Relaxation for Interpretable and Adaptive Network Control}

\author{Zhiyuan~Ren,~\IEEEmembership{Member,~IEEE},
        Zhiliang~Shuai,
        and~Wenchi~Cheng,~\IEEEmembership{Senior~Member,~IEEE},

\thanks{This work was supported by the National Key Research and Development Program of China under Grant No. 2024YFE0200300.}%
\thanks{Zhiyuan Ren, Zhiliang Shuai, and Wenchi Cheng are with the School of Telecommunications Engineering, Xidian University, Xi'an 710071, China.}%
\thanks{Corresponding author: Zhiyuan Ren (zyren@xidian.edu.cn).}%
}


\maketitle

\begin{abstract}
Prevailing network control strategies, which rely on static shortest-path logic, suffer from catastrophic "stress concentration" on critical nodes. This paper introduces the System Relaxation Algorithm (SRA), a new control paradigm inspired by physical relaxation that guides a network toward an emergent equilibrium of load balance. SRA is an interpretable, 'white-box' dynamical system whose behavior is profoundly topology-dependent: in heterogeneous networks, it acts as a proactive performance optimizer, reducing peak centrality by over 80\% and increasing high-load throughput by more than 45\%; in homogeneous topologies, its objective intelligently shifts to resilience enhancement. We rigorously prove its global convergence and practical stability using the theory of non-smooth dynamical systems, establishing a predictable paradigm for network governance that intelligently trades off performance and resilience.
\end{abstract}


\section{Introduction}
\label{sec:introduction}

\IEEEPARstart{P}{revailing} network control paradigms, rooted in static shortest-path logic, inherently lack a mechanism for managing structural stress. This oversight leads to "stress concentration," a severe imbalance of load on a few critical nodes. We formalize this "stress" using the network science metric of \textbf{Betweenness Centrality (BC)}, where a high BC value identifies a node as a topological bottleneck and a point of structural vulnerability. This endemic issue manifests as two critical symptoms: structurally, it creates "stress singularities" that make the network fragile to cascading failures; performatively, it causes premature capacity saturation, where bottlenecks lead to massive packet loss and soaring latency long before the network's physical capacity is reached.

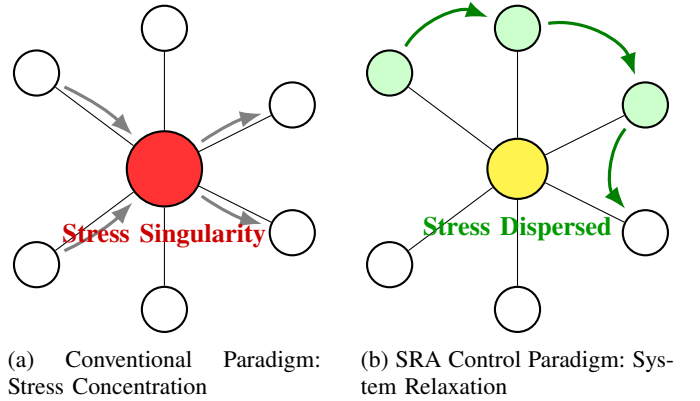
\begin{figure}[t!]
    \centering
    \begin{subfigure}[b]{0.47\columnwidth} 
        \centering
        \begin{tikzpicture}[scale=0.85] 
            \tikzstyle{node}=[circle, draw, thick, minimum size=6mm]
            \tikzstyle{hub}=[node, fill=red!80, minimum size=10mm]
            \tikzstyle{flow}=[-Latex, gray, thick, shorten <=1mm, shorten >=1mm]
            \node[hub] (H) at (0,0) {};
            \node[node] (N1) at (2,1) {};
            \node[node] (N2) at (2,-1) {};
            \node[node] (N3) at (-2,1.5) {};
            \node[node] (N4) at (-2,-1.5) {};
            \node[node] (P1) at (0, 2.2) {};
            \node[node] (P2) at (0, -2.2) {};
            \draw (H) -- (N1); \draw (H) -- (N2);
            \draw (H) -- (N3); \draw (H) -- (N4);
            \draw (H) -- (P1); \draw (H) -- (P2);
            \draw[flow, line width=1.2pt] (N3) to [bend left=10] (H);
            \draw[flow, line width=1.2pt] (N4) to [bend right=10] (H);
            \draw[flow, line width=1.2pt] (H) to [bend left=10] (N1);
            \draw[flow, line width=1.2pt] (H) to [bend right=10] (N2);
            \node[below=0.1cm of H, text=red!80!black, font=\bfseries] {Stress Singularity};
        \end{tikzpicture}
        \caption{Conventional Paradigm: Stress Concentration}
        \label{fig:intro_before}
    \end{subfigure}
    \hfill
    \begin{subfigure}[b]{0.47\columnwidth} 
        \centering
        \begin{tikzpicture}[scale=0.85] 
            \tikzstyle{node}=[circle, draw, thick, minimum size=6mm]
            \tikzstyle{hub}=[node, fill=yellow!80, minimum size=8mm]
            \tikzstyle{periph}=[node, fill=green!20]
            \tikzstyle{flow}=[-Latex, green!50!black, thick, shorten <=1mm, shorten >=1mm]
            \node[hub] (H) at (0,0) {};
            \node[periph] (N1) at (2,1) {};
            \node[node] (N2) at (2,-1) {};
            \node[periph] (N3) at (-2,1.5) {};
            \node[node] (N4) at (-2,-1.5) {};
            \node[periph] (P1) at (0, 2.2) {};
            \node[node] (P2) at (0, -2.2) {};
            \draw (H) -- (N1); \draw (H) -- (N2);
            \draw (H) -- (N3); \draw (H) -- (N4);
            \draw (H) -- (P1); \draw (H) -- (P2);
            \draw[flow, line width=1.2pt] (N3) to [bend left=40] (P1);
            \draw[flow, line width=1.2pt] (P1) to [bend left=40] (N1);
            \draw[flow, line width=1.2pt] (N1) to [bend right=40] (N2);
            \node[below=0.1cm of H, text=green!60!black, font=\bfseries] {Stress Dispersed};
        \end{tikzpicture}
        \caption{SRA Control Paradigm: System Relaxation}
        \label{fig:intro_after}
    \end{subfigure}
    
    \caption{Conceptual illustration of the core problem and the proposed SRA paradigm. SRA mitigates the "stress concentration" inherent in conventional network control by guiding the system to a "relaxed" state of balanced load.}
    \vspace{-4mm}
    \label{fig:intro_scenario}
\end{figure}

The root of this fragility lies in a flawed focus. Existing methods, including adaptive routing that reacts to transient metrics like queue length, concentrate on \textit{pathfinding} within a given cost landscape. They fail to address the underlying structural cause because they do not attempt to \textit{control the landscape itself}. This paper proposes a paradigm shift from reactive pathfinding to proactive stress control, inspired by the concept of "relaxation" in physical systems. We introduce the System Relaxation Algorithm (SRA), a model-driven control system designed to proactively disperse this stress, guiding the network toward a more robust and "relaxed" state of balanced load, as conceptually illustrated in Fig.~\ref{fig:intro_scenario}. This control paradigm is perfectly aligned with modern architectures like Software-Defined Networking (SDN), where a central controller possesses the global view required for stress calculation and the direct mechanisms to enforce the resulting cost policies.

Our main contributions are:
\begin{enumerate}
    \item \textbf{A Novel Control-Theoretic Model:} We formalize structural stress, quantified via betweenness centrality, as a controllable state variable within an interpretable, "white-box" dynamical system, introducing "system relaxation" as a new paradigm for network governance.
    \item \textbf{Rigorous Analysis of Discontinuous Dynamics:} We are the first to model the discrete path switching in network control as a non-smooth dynamical system, providing formal proofs of stability using specialized tools like piecewise Lyapunov functions.
    \item \textbf{Demonstrable Enhancement of Structural Robustness:} We demonstrate through simulations that SRA guides the network to self-organize into a verifiably more robust state, achieving an intelligent, topology-aware trade-off between performance and resilience.
\end{enumerate}

The remainder of this paper is organized as follows. Section~\ref{sec:related_work} reviews related work. Section~\ref{sec:model} details our theoretical model. Section~\ref{sec:simulation} presents simulation results, and Section~\ref{sec:discussion_conclusion} concludes the paper.

\section{Related Work}
\label{sec:related_work}

The challenge of routing in dynamic networks has spurred research beyond traditional protocols, which struggle to handle the "stress concentration" problem in complex topologies. This has led to two primary advanced methodologies: reactive, load-aware routing and adaptive, AI-driven routing. In contrast, the System Relaxation Algorithm (SRA) proposed here represents a third, model-driven paradigm that maintains full interpretability. Table I provides a comparative positioning of SRA against these approaches.

\begin{table}[h]
\centering
\caption{Comparative Positioning of SRA against Existing Routing Paradigms}
\label{tab:related_work_comparison}
\renewcommand{\arraystretch}{1.2}
\small
\begin{tabularx}{\columnwidth}{|l|X|X|X|}
\hline
\textbf{Feature} & \textbf{Traditional Load-Aware} & \textbf{AI-Driven} & \textbf{SRA (This Work)} \\
\hline \hline
\textbf{Sensing Object} & Real-time traffic state (e.g., queue lengths) & Traffic patterns \& Topology features & \textbf{Accumulated structural stress} \\
\hline
\textbf{Timescale} & Instantaneous / Short-term & Short to Medium-term & \textbf{Long-term structural equilibrium} \\
\hline
Primary Goal & Mitigate local, transient congestion & Maximize performance metrics (e.g., throughput) & Disperse endemic structural stress \\
\hline
Core Method & Heuristic cost functions & Black-box learning (e.g., RL, GNN) & Model-driven dynamical system \\
\hline
Interpretability & High & Low & High \\
\hline
\end{tabularx}
\end{table}

\subsection{Traditional Load-Aware Routing}
Load-aware routing protocols dynamically adjust paths based on real-time network state metrics. These metrics often include local information such as node queue lengths \cite{14}, channel busyness \cite{11}, or composite path load metrics \cite{4}. For instance, the ALR protocol proposed by S and M \cite{11} uses a composite metric of queue length and packet transmission success rate to select paths. Similarly, the CRADLE protocol by Macabale et al. \cite{14} leverages queue occupancy and channel state information for path selection in wireless mesh networks. The LIRP protocol by Tao et al. \cite{4} employs a cost function based on link quality and queue length. While effective at mitigating immediate, localized congestion, these methods often rely on heuristic rules and lack a global perspective on network structure. They primarily focus on state-aware path selection rather than fundamentally reshaping the underlying "cost landscape" of the network based on its structural properties, a key differentiator of our SRA framework.

\subsection{Static Centrality-Based Routing Optimization}
Centrality metrics, which quantify the structural importance of nodes, have been widely used to inform routing decisions. While some studies have explored physical layering \cite{1} or closeness centrality \cite{13}, betweenness centrality (BC) \cite{9, 12} is the most common metric for identifying structural bottlenecks. Several studies have proposed static or quasi-static methods that leverage BC. For example, Ahmed et al. \cite{7} and Ma and Yang \cite{18} used pre-calculated BC to optimize routing paths and flow table storage, respectively. Upadhyay and Singh \cite{6} attempted to use a genetic algorithm to find a single, static network configuration that minimizes the variance of BC. However, calculating BC is computationally intensive \cite{8}, and recent efforts have focused on distributed computation methods \cite{17, 19}. The primary limitation of these approaches is their static or infrequently updated nature, which struggles to adapt to the rapid, dynamic changes in network topology and traffic that our SRA framework is designed to handle.

\subsection{AI-Driven Adaptive Routing}
Modern AI techniques, particularly machine learning and reinforcement learning, have been applied to create highly adaptive routing protocols. These methods can be broadly categorized as either heuristic optimization or reinforcement learning-based. Heuristic approaches like Genetic Algorithms (GA) \cite{2} and Particle Swarm Optimization (PSO) \cite{10, 16} can search for optimal routing policies but often require significant computational resources.

Reinforcement learning (RL) offers a more online, adaptive approach. Classical Q-learning algorithms \cite{15} have been used to allow nodes to learn optimal next-hops based on local rewards. More advanced methods now combine Graph Neural Networks (GNNs) with Deep Reinforcement Learning (DRL) \cite{3, 5} to learn complex relationships between network topology and traffic state for more sophisticated path selection. While powerful, these AI-driven methods often function as "black boxes," making their behavior difficult to predict, interpret, or formally verify. This lack of interpretability is a significant barrier to their deployment in mission-critical systems. In contrast, SRA is a fully model-driven, "white-box" framework whose behavior is governed by a clear, mathematically tractable dynamical system, a key advantage for mission-critical applications.

\subsection{Positioning of Our Work}
\label{subsec:positioning}

The novelty of the SRA framework lies in its unique synthesis of ideas from prior routes to create a system that is simultaneously: 1) \textbf{structural and proactive}, unlike symptomatic and reactive load-aware methods; 2) \textbf{dynamic and closed-loop}, unlike static centrality optimization; and 3) \textbf{mechanistically interpretable}, unlike "black-box" AI models. These fundamental distinctions are summarized in Table~\ref{tab:related_work_comparison}. SRA's core contribution, therefore, is its systematic, cross-disciplinary integration of three domains: employing centrality measures from \textbf{network science} as the sensing signal, utilizing an integrator from \textbf{control theory} as the actuation engine, and applying the theory of \textbf{non-smooth dynamical systems} to provide rigorous mathematical guarantees for its stable operation. It is this integration that establishes a new, model-driven paradigm for resilient network design.

\section{Theoretical Model and Methods}
\label{sec:model}

The primary challenge in designing any adaptive routing protocol is ensuring provable stability, as naive reactive approaches often collapse into destructive oscillations. To overcome this, we move beyond simple heuristics and formally construct SRA as a \textbf{discrete-time, non-smooth dynamical system}. This control-theoretic approach is essential for modeling the complex interplay between the continuous evolution of node costs and the discrete "switching" of routing paths. This section provides a systematic exposition of this system: we first define the network model and the centrality-based state variables, then detail the core leaky-integrator evolution rule. The culmination is a rigorous proof of SRA's global convergence and practical stability using the powerful tools of piecewise Lyapunov functions and discrete-time inclusion theory. 

\subsection{Network Model and Load Metric}

We model the network as an undirected graph $G=(V,E)$, where $N=|V|$. Throughout SRA's iterative process, path lengths are calculated based on a dynamically evolving node cost vector $C \in \mathbb{R}^N_+$. The cost of any path $P$ is the sum of the transit costs of its constituent nodes, i.e., $d(P) = \sum_{v \in P} C_v$. Standard shortest-path algorithms are used to minimize this cumulative cost. We assume $G$ is connected and $N \ge 3$.

To quantify the load concentration induced by the topology and cost distribution, we employ \textbf{Betweenness Centrality (BC)} as the core metric. The (unnormalized) BC of a node $i$, denoted $\kappa_i(S)$, is the total number of all-pairs shortest paths that pass through it:
\begin{equation}
    \label{eq:bc_definition}
    \kappa_i(S) = \sum_{s\neq i\neq t\in V} \frac{\sigma^{\,C(S)}_{st}(i)}{\sigma^{\,C(S)}_{st}}
\end{equation}
where $\sigma^{\,C(S)}_{st}$ is the number of shortest paths from $s$ to $t$, and $\sigma^{\,C(S)}_{st}(i)$ is the subset of those paths traversing $i$. As $\kappa_i(S)$ sums proportions over all node pairs, it is an unbounded real number reflecting a node's absolute bridging role. To ensure the boundedness of state variables in our dynamical system, we use a max-normalized centrality vector in the algorithm's update steps:
\begin{equation}
    \label{eq:bc_normalization}
    \mathrm{norm}(\kappa(S)) :=
    \begin{cases}
    \displaystyle \frac{\kappa(S)}{\max_{j}\kappa_j(S)}, & \text{if } \max_j \kappa_j(S) > 0,\\[6pt]
    \mathbf{0}, & \text{otherwise.}
    \end{cases}
\end{equation}
Consequently, the vector $\mathrm{norm}(\kappa(S))$ is always confined within the unit hypercube $[0,1]^N$. Crucially, for a finite graph with a deterministic tie-breaking rule, the set of all possible shortest-path configurations is finite. This implies that the set of all possible values for $\mathrm{norm}(\kappa(S))$ is also a \textbf{finite set of constant vectors}. This property is a cornerstone of the subsequent convergence analysis.

\subsection{SRA: Core Mechanism and Algorithm Implementation}
\label{subsec:sra_mechanism}

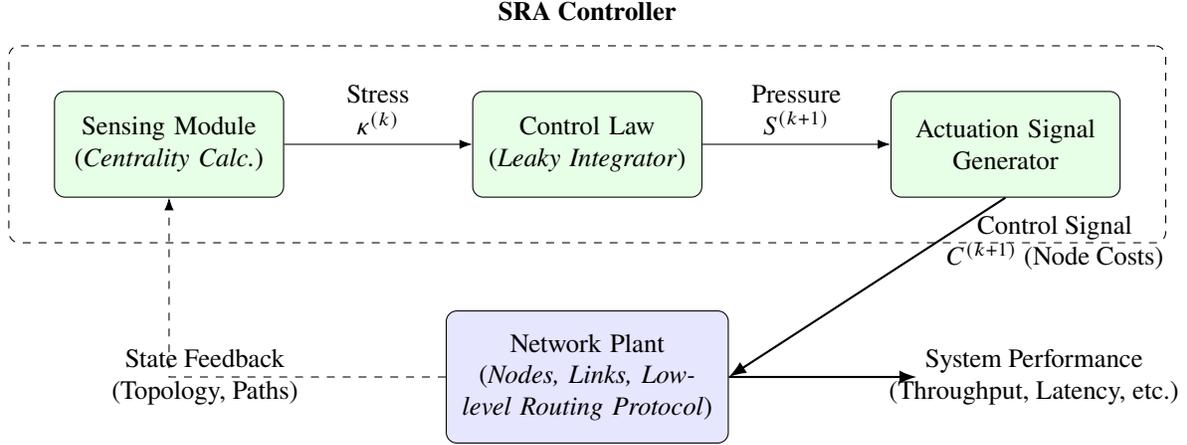
\begin{figure*}[h!]
    \centering
    \begin{tikzpicture}[
        auto,
        node distance=1.5cm and 2.5cm,
        plant/.style={rectangle, draw, fill=blue!10, text width=10em, text centered, rounded corners, minimum height=5em},
        module/.style={rectangle, draw, fill=green!10, rounded corners, text width=8em, text centered, minimum height=4em},
        line/.style={draw, -{Latex}},
        signal/.style={draw, -{Latex}, thick},
        feedback/.style={draw, -{Latex}, dashed}
    ]
    \node [module] (sensing) {Sensing Module \\ (\textit{Centrality Calc.})};
    \node [module, right=of sensing] (law) {Control Law \\ (\textit{Leaky Integrator})};
    \node [module, right=of law] (actuation) {Actuation Signal Generator};
    \node [plant, below=of law] (plant) {Network Plant \\ (\textit{Nodes, Links, Low-level Routing Protocol})};
    \node [draw, rounded corners, fit=(sensing) (law) (actuation), label={[yshift=0.2cm]above:\textbf{SRA Controller}}, inner sep=0.6cm, dashed] (controller_box) {};
    \path [signal] (actuation.south) -- node [right, pos=0.25, align=center] {Control Signal \\ $C^{(k+1)}$ (Node Costs)} (plant.east);
    \path [feedback] (plant.west) -| node [left, pos=0.25, align=center] {State Feedback \\ (Topology, Paths)} (sensing.south);
    \path [line] (sensing.east) -- node [above, align=center] {Stress \\ $\kappa^{(k)}$} (law.west);
    \path [line] (law.east) -- node [above, align=center] {Pressure \\ $S^{(k+1)}$} (actuation.west);
    \coordinate[right=of plant] (output_coord);
    \path [signal] (plant.east) -- node [right, pos=0.8, align=center] {System Performance \\ (Throughput, Latency, etc.)} (output_coord);
    \end{tikzpicture}
    \caption{The architecture of the SRA as a closed-loop control system. The SRA controller acts as a supervisory layer that governs the underlying network plant by dynamically adjusting the cost landscape.}
    \label{fig:sra_architecture}
\end{figure*}

Having defined the network model and load metric, this section details SRA's dynamic cost evolution rule. We formally design SRA as a closed-loop control system, as depicted in Fig.~\ref{fig:sra_architecture}, where the SRA controller governs the underlying network plant.

A naive feedback mechanism making node cost proportional to its current centrality is inherently unstable and leads to persistent oscillations. To ensure stability, SRA's design must incorporate memory. We achieve this using an Integral (I) control structure, implemented as a \textbf{leaky integrator}. A core state vector, which we term the "accumulated pressure" $S \in \mathbb{R}^N_+$, represents the historically accumulated stress on each node. This state evolves according to the discrete-time transition equation:
\begin{equation}
    \label{eq:pressure_update}
    S_i^{(k+1)} = (1 - \alpha) S_i^{(k)} + \alpha \cdot \mathrm{norm}(\kappa_i^{(k)})
\end{equation}
where $k$ is the iteration step and the learning rate $\alpha \in (0,1)$ controls the system's memory. The $(1-\alpha)$ term provides a critical \textbf{damping} effect, ensuring the stability of the integration process. The node transit cost $C$—the system's control output—is then defined as an affine function of this historical state:
\begin{equation}
    \label{eq:cost_update}
    C_i^{(k+1)} = 1 + \beta_I \cdot S_i^{(k+1)}
\end{equation}
where $1$ is a base cost and $\beta_I > 0$ is the "pressure-to-cost" conversion factor. By basing costs on the system's history ($S$), SRA makes stable, proactive decisions.

\begin{figure}[hbtp]
    \centering
    \includegraphics[width=0.9\columnwidth]{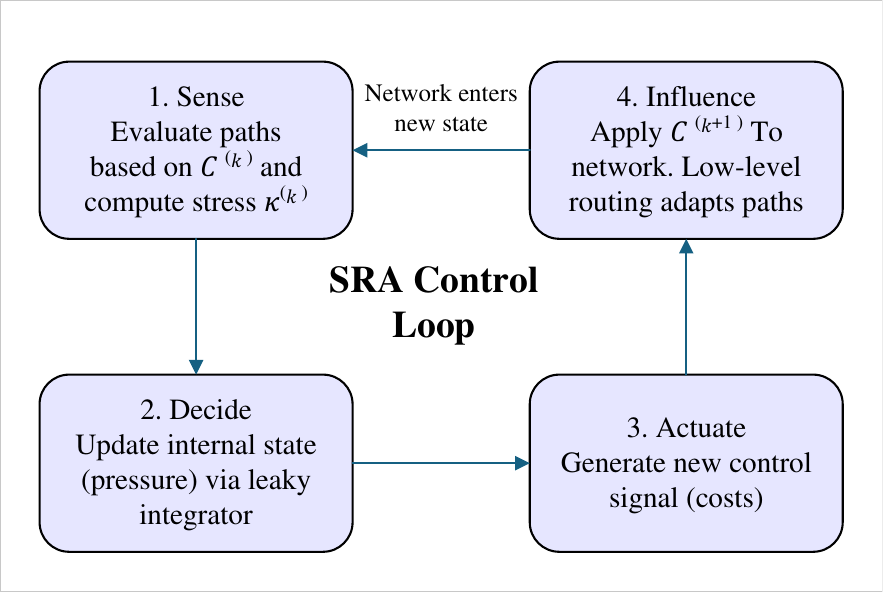}
    \caption{The iterative workflow of the SRA control loop, illustrating the four key phases of sensing, decision-making, actuation, and system influence that drive the network toward a state of dispersed structural stress.}
    \label{fig:sra_workflow}
\end{figure}

\begin{figure}[t!]
    \centering
    \includegraphics[width=0.8\columnwidth]{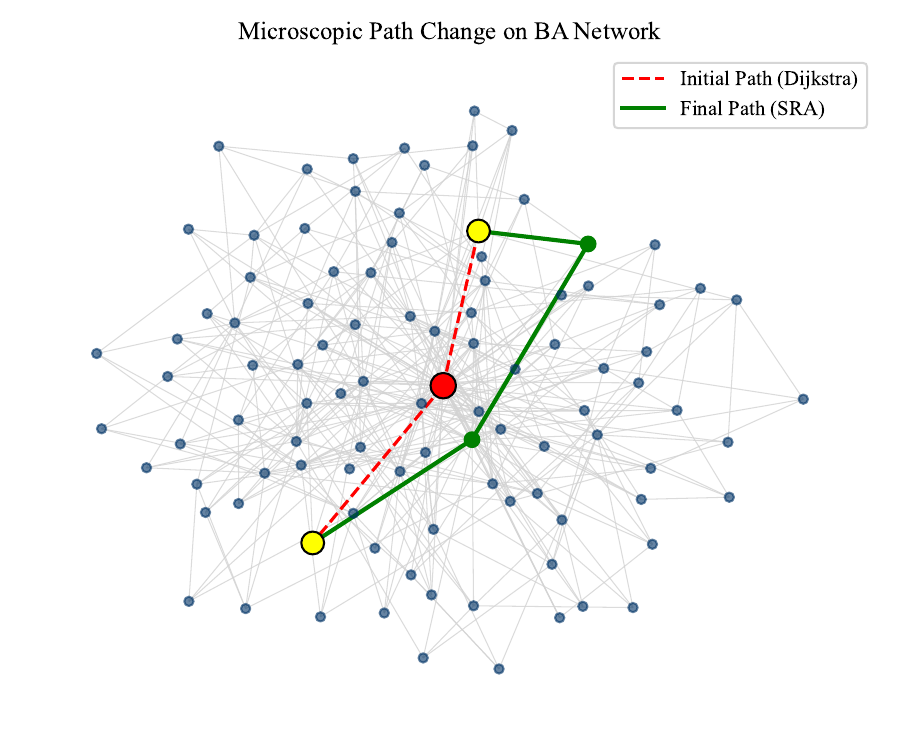}
    \caption{A conceptual visualization of SRA's path reshaping mechanism. The algorithm diverts traffic from the efficient but congested 2-hop path (red dashed) through the central hub to a longer but more stable 4-hop peripheral path (green solid).}
    \label{fig:sra_mechanism_visualization}
\end{figure}

The complete iterative process is described in Algorithm~\ref{alg:sra}, and its cyclical, four-phase workflow is visualized in Fig.~\ref{fig:sra_workflow}. The algorithm introduces a key tunable parameter, the minimum dwell time $T_d$, which allows network operators to trade off path stability against system responsiveness. The core principle of SRA is to reshape the network's "cost landscape" based on accumulated structural stress. As shown conceptually in Fig.~\ref{fig:sra_mechanism_visualization}, this forces the routing protocol to intelligently trade path efficiency for bottleneck avoidance, rerouting traffic from a physically short but high-stress path to a longer but more stable peripheral route.

\begin{algorithm}[h!]
    \DontPrintSemicolon
    \KwIn{Graph $G=(V,E)$; learning rate $\alpha \in (0,1)$; pressure-cost factor $\beta_I > 0$; stopping threshold $\varepsilon > 0$; max iterations $K_{\max}$; (optional) min dwell time $T_d \ge 1$.}
    \KwOut{Final node cost vector $C^{(k+1)}$ and the corresponding routing policy.}
    \BlankLine
    \textbf{Initialize:} $S^{(0)} \leftarrow \mathbf{0}$; $C^{(0)} \leftarrow \mathbf{1}$\;
    \BlankLine
    \For{$k=0, 1, 2, \dots, K_{\max}-1$}{
        \tcp{Step 1: Path Evaluation}
        Compute all-pairs shortest paths using node costs $C^{(k)}$\;
        \tcp{Step 2: Centrality Calculation}
        Compute unnormalized centrality $\kappa^{(k)}$ based on new paths\;
        \tcp{Step 3: Pressure Update}
        $S^{(k+1)} \leftarrow (1-\alpha)S^{(k)}+\alpha\,\mathrm{norm}(\kappa^{(k)})$\;
        \tcp{Step 4: Cost Update}
        $C^{(k+1)} \leftarrow 1+\beta_I\,S^{(k+1)}$\;
        \tcp{Step 5: Anti-Chattering Strategy}
        \If{a path set switch occurred at $k$ and $T_d > 1$}{
            Forcibly use current path set for the next $T_d-1$ steps, executing only steps 2-6\;
        }
        \tcp{Step 6: Convergence Check}
        \If{$\|S^{(k+1)}-S^{(k)}\|_\infty \le \varepsilon$ and path set is stable}{
            \textbf{break}\;
        }
    }
    \caption{System Relaxation Algorithm (SRA)}
    \label{alg:sra}
\end{algorithm}

The computational complexity of a single SRA iteration is dominated by the all-pairs shortest-path and BC calculation, which is $O(N(E + N \log N))$ using Brandes' algorithm. While this cost can be substantial for large-scale networks, it is computationally feasible in the intended semi-dynamic or SDN-based deployment scenarios where a central entity performs the calculation. Furthermore, the centrality update is highly parallelizable, and future work can integrate the SRA framework with high-performance approximation algorithms, creating an effective trade-off between optimization quality and overhead for fully dynamic environments.

\subsection{Theoretical Analysis of Model Convergence}
\label{sec:convergence_analysis}

This section provides a rigorous analysis of SRA's dynamical behavior. The inherent, discontinuous switching of routing paths makes the system \textbf{non-smooth}, requiring specialized analytical tools. Our analysis is structured as follows: first, we prove the existence of a global attractor for the system's discrete-time inclusion model, guaranteeing its trajectories are ultimately bounded. Second, we use a piecewise Lyapunov function to quantitatively characterize its oscillatory behavior. Finally, we introduce a minimum dwell time as an engineering mechanism to achieve practical stability and derive an explicit convergence rate.

\subsubsection{Global Attractiveness and Existence Guarantee for Convergence}
\label{subsec:attractivity}

The convergence analysis of SRA is challenged by the piecewise constant, non-smooth nature of the centrality map $\kappa(S)$. To establish a rigorous guarantee, we reformulate the algorithm as a \textbf{discrete-time inclusion}, a standard tool from non-smooth dynamical systems. The update rule is regularized into the set-valued map:
\begin{equation}\label{eq:inclusion_map}
S^{(k+1)} \in F(S^{(k)}) := (1-\alpha)S^{(k)}+\alpha\,\mathcal{N}(S^{(k)})
\end{equation}
where $\mathcal{N}(S)$ is a Krasovskii regularization of the discontinuous mapping $\mathrm{norm}(\kappa(\cdot))$, which takes the convex hull of all limit points in a vanishing neighborhood of $S$. This yields a mathematically well-behaved map, allowing us to prove convergence.

\begin{lemma}[Positive Invariance and Absorbing Set]
\label{lemma:invariance}
Let the state space be $X=[0,1]^N$. The discrete-time inclusion for SRA, given by $S^{(k+1)}\in F(S^{(k)})$, evolves on a compact and positively invariant set $X$, which also serves as a compact absorbing set for the system.
\end{lemma}
\begin{IEEEproof}[Proof Sketch]
The proof shows that if a state $S^{(k)}$ is within the unit hypercube $X$, its successor $S^{(k+1)}$—being a convex combination of elements in $X$—must also lie within $X$. This establishes positive invariance, from which the absorbing property follows. \textit{The full formal proof is provided in Appendix~\ref{app:proof_lemma_invariance}.}
\end{IEEEproof}

Lemma~\ref{lemma:invariance} provides a foundational safety guarantee against divergence. Next, we verify the map's analytical properties.

\begin{proposition}[Upper Semi-Continuity and Compact, Convex Values]
\label{prop:usc}
The set-valued map $F(\cdot)$ defined in \eqref{eq:inclusion_map} is an \textbf{upper semi-continuous (USC)} map with \textbf{non-empty, compact, and convex} values.
\end{proposition}
\begin{IEEEproof}[Proof Sketch]
The proof relies on the piecewise constant structure of $\mathcal{N}(S)$, which stems from the finite nature of the network's path configurations. This structure ensures that the graph of the map is closed, which in turn guarantees the USC property. \textit{The full formal proof is provided in Appendix~\ref{app:proof_prop_usc}.}
\end{IEEEproof}

With these properties established, we can prove the system's global convergence to a stable equilibrium set.

\begin{theorem}[Existence of a Global Attractor]
\label{thm:attractor}
For the discrete-time inclusion \eqref{eq:inclusion_map}, there exists a non-empty, compact, and positively invariant set $\Omega\subset X:=[0,1]^N$ such that for any initial value $S^{(0)}\in X$,
\begin{equation}\label{eq:attractor_convergence}
\mathrm{dist}\big(S^{(k)},\Omega\big)\ \to\ 0\qquad (k\to\infty).
\end{equation}
\end{theorem}
\begin{IEEEproof}[Proof Sketch]
The proof is based on standard arguments for non-smooth dynamical systems. Since the system evolves on a compact, positively invariant set (Lemma~\ref{lemma:invariance}) and is governed by a USC map (Proposition~\ref{prop:usc}), any trajectory is pre-compact. The $\omega$-limit set of any such trajectory is then shown to be non-empty and invariant, allowing for the construction of a global attractor $\Omega$. \textit{The full formal proof is provided in Appendix~\ref{app:proof_thm_attractor}.}
\end{IEEEproof}

This theorem definitively guarantees that the SRA algorithm will always converge to a bounded, stable region $\Omega$ and will never diverge.

\subsubsection{Piecewise Lyapunov Analysis and Ultimate Boundedness}
\label{subsec:lyapunov}

The existence of a global attractor guarantees non-divergence, but does not quantitatively characterize the system's final dynamics, such as persistent oscillations. To rigorously answer whether the algorithm will "oscillate indefinitely" and to bound this behavior, this section introduces a powerful tool from the analysis of non-smooth switched systems: the \textbf{piecewise Lyapunov function}.

Our goal is to construct a function $V(S)$ that quantifies the system's "energy"—defined as its squared Euclidean distance to an ideal equilibrium set. We define it as:
\begin{equation}\label{eq:lyapunov_def}
V(S):=\min_{c\in\mathcal N(S)}\|S-c\|_2^2.
\end{equation}
The existence of this minimum is guaranteed as $\mathcal N(S)$ is non-empty and compact. Concurrently, we define $\bar{\Delta} := \max_{i,j} \|v_i - v_j\|$ as the diameter of the finite set of all possible constant centrality vectors $\{v_\ell\}$. The following theorem reveals the fundamental evolution law for this system "energy."

\begin{theorem}[Drop-Jump Inequality and Ultimate Boundedness]
\label{thm:drop_jump}
 For any step $k$, the Lyapunov function satisfies the inequality:
 \begin{equation}\label{eq:drop_jump_one_step}
 V(S^{(k+1)})\ \le\ (1-\alpha)\,V(S^{(k)})\ +\ \dfrac{1}{\alpha}\,\bar\Delta^{2}.
 \end{equation}
 Consequently, the system is ultimately bounded, with the limit superior of $V(S^{(k)})$ satisfying:
\begin{equation}\label{eq:limsup_boundedness}
\limsup_{k\to\infty}V(S^{(k)})\ \le\ \frac{\bar\Delta^{2}}{\alpha^2}.
\end{equation}
\end{theorem}
\begin{IEEEproof}[Proof Sketch]
The proof hinges on bounding the one-step evolution of $V(S)$. By applying the Weighted Young's Inequality, we establish a linear recurrence for $V(S^{(k)})$ that explicitly reveals its 'drop' and 'jump' terms, from which ultimate boundedness directly follows. \textit{The full formal proof is provided in Appendix~\ref{app:proof_thm_drop_jump}.}
\end{IEEEproof}

This "drop-jump" inequality is key to understanding SRA's stability. It shows that the system's energy $V(S)$ is governed by two competing forces: an energy \textbf{decay term} (the "drop") with a contraction factor of $(1-\alpha)$, representing the algorithm's intrinsic damping; and an energy \textbf{injection term} (the "jump"), upper-bounded by $\frac{1}{\alpha}\bar\Delta^2$ and caused by path switching. This theorem proves that SRA's oscillatory behavior is fully controllable. To better understand the damping effect, the following corollary analyzes the ideal case where no path switching occurs.

\begin{corollary}[Local Exponential Convergence in a Fixed Partition]
\label{cor:local_convergence}
If the set of shortest paths does not switch, the update rule becomes a standard linear contraction map:
\begin{equation}\label{eq:linear_update}
S^{(k+1)}=(1-\alpha)S^{(k)}+\alpha c,
\end{equation}
which converges exponentially to the unique fixed point $S^\star=c$ at a rate of $(1-\alpha)$.
\end{corollary}
\begin{IEEEproof}[Proof Sketch]
The proof first solves the fixed-point equation $S=(1-\alpha)S+\alpha c$ to show uniqueness of $S^\star=c$. It then subtracts the fixed point from the update rule to derive the linear contraction relationship $\|S^{(k+1)}-S^\star\|_2 = (1-\alpha)\|S^{(k)}-S^\star\|_2$, from which exponential convergence directly follows. \textit{The full formal proof is provided in Appendix~\ref{app:proof_cor_local_convergence}.}
\end{IEEEproof}

\subsubsection{Convergence Rate and Practical Stability with Dwell Time}
\label{subsec:practical_stability_rate}

The "drop-jump" inequality guarantees that the system is ultimately bounded. However, for practical engineering applications, we must answer two further quantitative questions: How fast does the system converge to this stable neighborhood, and can we eliminate the residual oscillations (chattering)? This section shows that introducing a \textbf{minimum dwell time, $T_d$}, provides a definitive positive answer to both. By decomposing the system's evolution into "dwell blocks," we can derive an explicit convergence rate and formally prove that this strategy ensures practical stability.

To analyze the system under this policy, we decompose its evolution into "dwell blocks." Let the minimum dwell time be $T_d$. We define the start of each block recursively: $\tau_0=0$ and $\tau_{m+1} = \tau_m + T_d$. The $m$-th dwell block is the time interval $[\tau_m, \tau_{m+1}-1]$. We denote the state and Lyapunov energy at the start of block $m$ as $S_m := S^{(\tau_m)}$ and $V_m := V(S^{(\tau_m)})$, respectively.

\begin{proposition}[Intra-Block Contraction and Inter-Block Recurrence]
\label{prop:block_dynamics}
Under the minimum dwell time $T_d$ policy, the set of shortest paths remains constant within a block. Let $\gamma:=1-\alpha$, $\tilde C:=\bar\Delta^2/\alpha$. The state and energy at the block endpoints are $S_m$ and $V_m$ respectively. Then,
\begin{itemize}
    \item[\textit{(i)}] (Intra-block contraction) The energy contracts exponentially within each block:
    \begin{equation}\label{eq:intra_block}
    V(S^{(\tau_m+t)})\ \le\ \gamma^{\,t}V_m, \quad \text{for } t \in [0, T_d-1].
    \end{equation}
    \item[\textit{(ii)}] (Inter-block recurrence) The energies at the start of adjacent blocks are related by the linear recurrence:
    \begin{equation}\label{eq:inter_block}
    V_{m+1}\ \le\ \gamma^{T_d}\,V_m\ +\ \tilde C.
    \end{equation}
\end{itemize}
\end{proposition}
\begin{IEEEproof}[Proof Sketch]
The proof establishes two points. (i) For intra-block contraction, since the path set is fixed, the "jump" term in the Lyapunov inequality is zero, leading to simple exponential decay. (ii) For the inter-block recurrence, we chain the result from (i) over $T_d-1$ steps of decay with a single "drop-jump" step at the block boundary to derive the stated linear recurrence for $V_m$. \textit{The full formal proof is provided in Appendix~\ref{app:proof_prop_block_dynamics}.}
\end{IEEEproof}

This recurrence relation is the key to the global rate analysis. By solving it, we obtain explicit upper bounds for the steady-state radius and the number of steps required to converge.

\begin{theorem}[Steady-State Radius and Step Bound to an $\varepsilon$-Tube]
\label{thm:epsilon_tube}
Let $\gamma=1-\alpha\in(0,1)$ and $\tilde C=\bar\Delta^2/\alpha$. Then,
\begin{itemize}
    \item[\textit{(a)}] (Steady-state radius) The block-endpoint energy is ultimately bounded by:
    \begin{equation}\label{eq:limsup_radius}
    \limsup_{m\to\infty}V_m\ \le\ \frac{\tilde C}{1-\gamma^{T_d}}.
    \end{equation}
    \item[\textit{(b)}] (Steps to $\varepsilon$-tube) For a target radius $\varepsilon^2 > \frac{\tilde C}{1-\gamma^{T_d}}$, the number of steps $k$ to reach the set $\{S: V(S)\le \varepsilon^2\}$ is bounded.
\end{itemize}
\end{theorem}
\begin{IEEEproof}[Proof Sketch]
The proof is based on solving the linear recurrence relation derived in Proposition 2. By iteratively expanding the recurrence, we obtain a closed-form upper bound on $V_m$ as a geometric series, from which the limit superior (steady-state radius) and the number of steps to reach an $\varepsilon$-tube can be directly calculated. \textit{The full formal proof is provided in Appendix~\ref{app:proof_thm_epsilon_tube}.}
\end{IEEEproof}

Finally, the following theorem consolidates all benefits of the dwell-time policy into a single, powerful guarantee of practical stability.

\begin{theorem}[Dwell Time $\Rightarrow$ Practical Stability]
\label{thm:practical_stability}
With a minimum dwell time $T_d \ge 1$, the following conclusions hold:
\begin{itemize}
    \item[\textit{(a)}] \textbf{Finite Switching / No Chattering:} The number of path set switches $N_s(T)$ within any time window of length $T$ is bounded by $N_s(T)\ \le\ 1+\lfloor T/T_d \rfloor$.
    \item[\textit{(b)}] \textbf{Practical Stability:} The system trajectories are ultimately bounded within a neighborhood whose size is explicitly determined by $T_d$, as given in Theorem~\ref{thm:epsilon_tube}.
    \item[\textit{(c)}] \textbf{Convergence Time:} An explicit upper bound on the time to reach this neighborhood can be calculated as shown in Theorem~\ref{thm:epsilon_tube}.
\end{itemize}
\end{theorem}
\begin{IEEEproof}[Proof Sketch]
The proof consolidates the consequences of the dwell-time policy. (a) Finite switching is a direct result of the policy's definition. (b) Practical stability is a direct consequence of the steady-state radius calculated in Theorem~\ref{thm:epsilon_tube}. (c) The convergence time bound is obtained by inverting the inequality derived in Theorem 3. \textit{The full formal proof is provided in Appendix~\ref{app:proof_thm_practical_stability}.}
\end{IEEEproof}

This theoretical result provides a valuable design principle. Given a desired steady-state energy radius $R_{\text{target}}^2$, the minimum required dwell time $T_d$ to guarantee it can be determined:
\begin{equation}\label{eq:design_parameter_Td}
\frac{\tilde C}{1-\gamma^{T_d}}\ \le\ R_{\text{target}}^2
\quad\Longrightarrow\quad
T_d\ \ge\ \Big\lceil \frac{\ln(1-\tilde C/R_{\text{target}}^2)}{\ln \gamma}\Big\rceil.
\end{equation}
This clearly demonstrates how to quantitatively trade off system convergence speed against its final stability precision by tuning the single parameter $T_d$.

\subsection{Betweenness Peaks and Network Performance}
\label{subsec:theory_bridge}

The convergence guarantees established previously do not formally imply that the resulting equilibrium state offers improved network performance. This section provides this theoretical link, building a rigorous bridge from the algorithm's proven peak-shaving dynamics to concrete performance gains. We first prove that SRA guarantees a net reduction of the peak centrality.

\begin{lemma}[Blockwise Peak Drop]
\label{lem:block-budget}
Within a dwell-time block that includes a path-switching event, if the network has no structurally unavoidable bottlenecks, then the peak unnormalized centrality is guaranteed to decrease, i.e., there exists some $\delta>0$ such that $\kappa^{(m+1)}_{\max}\le \kappa^{(m)}_{\max}-\delta$.
\end{lemma}
\begin{IEEEproof}[Proof Sketch]
The existence of a net peak decrease is a direct consequence of the system's guaranteed evolution within a stable attractor (Theorems~\ref{thm:attractor}, \ref{thm:practical_stability}). The state update rule ensures that a node with peak centrality experiences the strongest pressure increase, forcing an eventual traffic diversion that strictly reduces its unnormalized centrality. \textit{The full formal proof is provided in Appendix~\ref{app:proof_lem_block_budget}.}
\end{IEEEproof}

This guaranteed reduction in peak centrality, a direct result of SRA's control law, translates directly into performance improvements. The following proposition quantifies this gain without relying on any idealized traffic redistribution assumptions.

\begin{proposition}[Quantifying the Performance Gain]
\label{prop:perf-bridge}
Consider a network where arrivals are proportional to centrality, $\Lambda_i \propto \kappa_i$. Let $\kappa'$ be the centrality vector after applying SRA, and suppose the peak is reduced, $\kappa'_{\max}<\kappa_{\max}$. Then:
\begin{enumerate}[(i)]
\item \emph{Unconditional robust lower bound.} The robust lower bound on network capacity, defined as $L(\kappa):=\mu_{\min}/\kappa_{\max}$, strictly increases, i.e., $L(\kappa')>L(\kappa)$.
\item \emph{Aligned-bottleneck capacity gain.} If the original bottleneck was at the peak node, the actual network capacity $\rho_{\max}(\kappa):=\min_{i} \mu_i/\kappa_i$ strictly increases.
\end{enumerate}
\end{proposition}
\begin{IEEEproof}[Proof Sketch]
The proof leverages the guaranteed peak reduction from Lemma~\ref{lem:block-budget}. (i) The increase of the robust lower bound $L(\kappa)$ follows directly from its definition, as $\kappa'_{\max}$ strictly decreases while $\mu_{\min}$ is constant. (ii) For the aligned-bottleneck case, the new capacity $\rho_{\max}(\kappa')$ is lower-bounded by the post-SRA capacity of the original peak node, which is strictly greater than the original network capacity. \textit{The full formal proof is provided in Appendix~\ref{app:proof_prop_perf_bridge}.}
\end{IEEEproof}

\begin{remark}[Generality and Limitations]
A key strength of these performance guarantees is their generality; the proofs do not rely on simplifying assumptions regarding traffic redistribution and thus hold for the arbitrary, non-proportional shifts that occur in practice. However, if the topology contains unavoidable cuts or bridges, the ultimate capacity is limited by these structural features. In such cases, SRA still suppresses all \emph{non-structural} peaks, and the robust lower bound guarantee of Proposition~\ref{prop:perf-bridge}(i) continues to hold.
\end{remark}

\section{Simulation Analysis}
\label{sec:simulation}

The theoretical framework established in Section~\ref{sec:model} guarantees SRA's stability and links its peak-shaving mechanism to performance enhancement. This section provides comprehensive empirical evidence to validate these claims across diverse network conditions. We are guided by three objectives: first, to directly validate the core theoretical link between structural optimization and capacity gains; second, to analyze the macroscopic, topology-dependent dynamics of the "system relaxation" process; and finally, to benchmark SRA's performance in terms of throughput, reliability, and robustness against standard and adaptive routing protocols.

\subsection{Simulation Environment Setup}
\label{subsec:sim_setup}
\begin{table}[h!]
\centering
\caption{Summary of Simulation Parameters. Note on capacity: "Node Service Rate" ($\mu_i$) is the steady-state rate perceived by flows, set from the "Node Proc. Capacity" (hardware-level budget) after accounting for overheads.}
\label{tab:sim_params}
\renewcommand{\arraystretch}{1.2}
\small
\begin{tabularx}{\columnwidth}{|l|X|}
\hline
\textbf{Parameter} & \textbf{Value / Description} \\
\hline
\hline
\multicolumn{2}{|c|}{\textbf{A. Network Topology Parameters}} \\
\hline
Network Types & Barabási-Albert (BA), Watts-Strogatz (WS), Erd\H{o}s-Rényi (ER) \\
\hline
Default Size ($N$) & 100 nodes \\
\hline
BA Attachment ($m$) & 4 \\
\hline
WS Neighbors ($k$) & 8 \\
\hline
WS Rewiring ($p$) & 0.1 \\
\hline
ER Connection ($p$) & 0.1 \\
\hline
\hline
\multicolumn{2}{|c|}{\textbf{B. SRA Algorithm Parameters}} \\
\hline
Learning Rate ($\alpha$) & 0.1 \\
\hline
Integral Gain ($\beta_I$) & 10.0 (Default) \\
\hline
Max Iterations ($K_{\max}$) & 200 \\
\hline
Convergence Tol. ($\varepsilon$) & $1 \times 10^{-5}$ \\
\hline
\hline
\multicolumn{2}{|c|}{\textbf{C. Dynamic Traffic Simulation Parameters}} \\
\hline
Node Service Rate ($\mu$) & 1.0 packet/timestep \\
\hline
Total Timesteps ($T$) & 2000 \\
\hline
Warmup Period & 300 timesteps \\
\hline
Node Proc. Capacity & 10 packets/timestep \\
\hline
Node Queue Capacity & 50 packets \\
\hline
\end{tabularx}
\end{table}

To evaluate SRA's universality, we conduct experiments on three canonical network models with distinct structural properties: Barabási-Albert (BA), Watts-Strogatz (WS), and Erd\H{o}s-Rényi (ER). The default parameters for network generation, the SRA algorithm, and the discrete-event traffic simulator are summarized in Table~\ref{tab:sim_params}. For benchmark protocols, Dijkstra and LIRP-D are deterministic; Q-Routing was configured with a learning rate of $\alpha=0.1$, discount factor $\gamma=0.9$, and exploration rate $\epsilon=0.3$. All presented results are the mean of 100 independent trials with distinct random seeds; error bands, where shown, depict $\pm1$ standard deviation. The SRA controller's cost landscape was updated at a fixed interval of 50 timesteps in dynamic simulations.

\subsection{Empirical Validation of Core Theoretical Claims}
\label{subsec:empirical_validation}

\begin{figure}[h]
    \centering
    \includegraphics[width=\columnwidth]{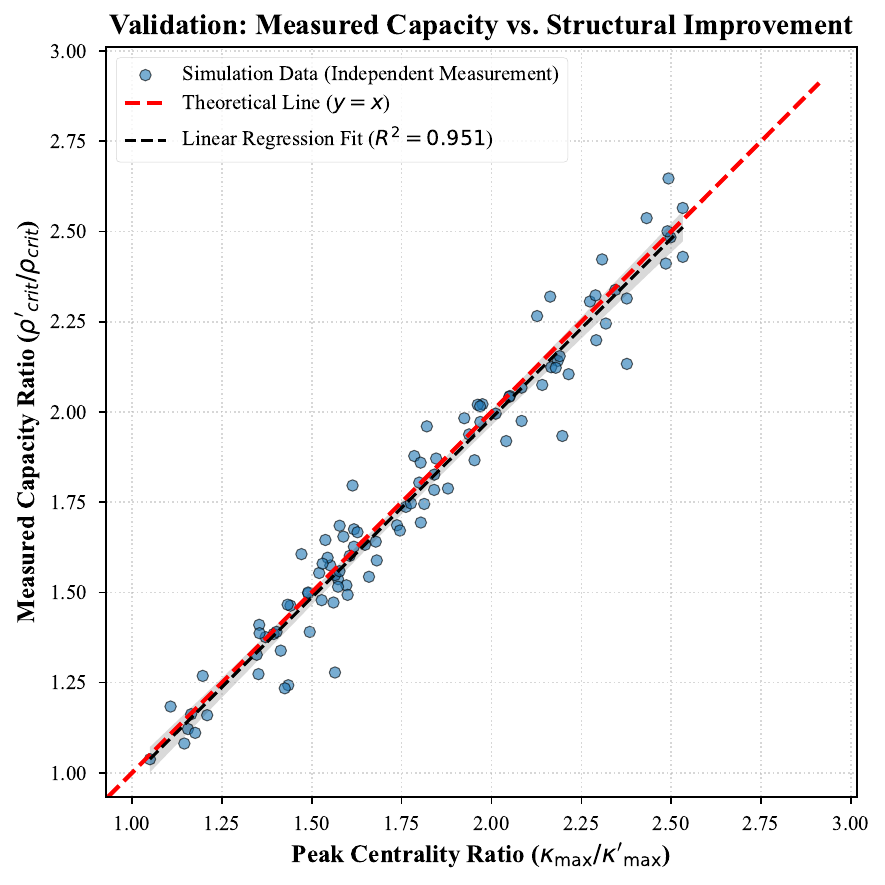}
    \caption{Independently Measured Network Capacity Gain vs. Structural Optimization Ratio. The tight clustering of empirical data around the theoretical $y=x$ line validates our core theory that performance gains are driven by structural stress dispersion. Sample details are in Section IV-A.}
    \label{fig:peak_validation}
\end{figure}

This section empirically validates the two central claims from our theoretical analysis: first, that SRA's structural optimization directly translates into quantifiable performance gains, and second, that the algorithm's non-smooth dynamics are stable and controllable as predicted.

To validate the structure-performance link, we conducted an experiment on 100 randomized networks (BA and WS topologies, $N \in [40, 60]$). We performed two independent measurements on each sample: a static analysis of the peak centrality reduction ratio ($\kappa_{\max}/\kappa'_{\max}$) achieved by SRA, and a dynamic simulation to measure the actual network capacity gain ($\rho'_{\text{crit}}/\rho_{\text{crit}}$). The results are shown in Fig.~\ref{fig:peak_validation}. Under the experimental condition of homogeneous node service rates ($\mu_i=\text{const}$), our theory (Proposition~\ref{prop:perf-bridge}(ii)) predicts a linear relation $\rho'_{\max}/\rho_{\max} = \kappa_{\max}/\kappa'_{\max}$. The tight clustering of empirical data around this $y=x$ line ($\mathbf{R^2=0.951}$) provides powerful validation, confirming that SRA's capacity enhancement is directly driven by its peak-shaving mechanism.

Furthermore, we validated the stability analysis presented in Section~\ref{sec:convergence_analysis}. Numerical experiments confirm that the evolution of the piecewise Lyapunov function $V(S)$ perfectly reproduces the "drop-jump" dynamics predicted by Theorem~\ref{thm:drop_jump}. Moreover, the introduction of a minimum dwell time ($T_d>1$) is shown to be a highly effective mechanism for eliminating the high-frequency "chattering" of the baseline algorithm, achieving the predictable, block-wise convergence guaranteed by Theorem~\ref{thm:practical_stability}. The detailed plots for these stability validations are provided in the Supplementary Material.

\subsection{Dynamic Relaxation of Structural Stress and Analysis of the Equilibrium State}
\label{subsec:dynamic_relaxation}

Having validated the underlying theory, we now analyze the structural properties of the equilibrium state that SRA achieves. To provide a statistically robust "before-and-after" comparison, we analyze the final network structure averaged over 100 independent runs. The results, summarized in Table~\ref{tab:structural_benefits_stats}, provide precise quantitative evidence of SRA's structural optimization benefits across different topologies.

The data highlight the algorithm's profound and topology-dependent impact. SRA's intervention is most aggressive in the heterogeneous BA network, which is prone to severe bottlenecks. Here, a minor cost—an average path length increase of only \textbf{5.63\%}—yields massive structural gains: a remarkable \textbf{83.76\%} reduction in peak centrality and a \textbf{75.38\%} decrease in the standard deviation of centrality. This is powerful evidence of "stress dispersion," where the burden on a few critical hubs is effectively redistributed across the network. In contrast, SRA's behavior in the more homogeneous WS and ER networks is more conservative. It still achieves significant structural improvements (e.g., ~49\% peak centrality reduction in WS networks) but with a negligible efficiency cost (less than 2\% for WS and none for ER).

These results offer strong quantitative support for our core thesis: SRA is a universal and highly effective control system for dispersing network structural stress, capable of applying a "targeted remedy" that intelligently trades a minor efficiency cost for a fundamental enhancement in structural stability.

\begin{table}[h!]
\centering
\caption{Analysis of Structural Optimization Benefits of SRA (Mean $\pm$ Std Dev over 100 runs)}
\label{tab:structural_benefits_stats}
\resizebox{\columnwidth}{!}{
\renewcommand{\arraystretch}{1.2}
\begin{tabular}{|l|l|r|r|r|}
\hline
\textbf{Network} & \textbf{Metric} & \textbf{Initial (Dijkstra)} & \textbf{Final (SRA)} & \textbf{Change (\%)} \\
\hline\hline
ER & Peak Centrality & $589.93 (\pm 120.59)$ & $209.40 (\pm 26.87)$ & -64.50\% \\
 & Std Dev of Centrality & $106.93 (\pm 10.62)$ & $33.55 (\pm 9.26)$ & -68.62\% \\
 & Avg Path Length & $2.25 (\pm 0.04)$ & $2.25 (\pm 0.04)$ & +0.00\% \\
\hline
BA & Peak Centrality & $2657.37 (\pm 619.87)$ & $431.52 (\pm 45.86)$ & -83.76\% \\
 & Std Dev of Centrality & $370.10 (\pm 43.87)$ & $91.12 (\pm 6.86)$ & -75.38\% \\
 & Avg Path Length & $2.37 (\pm 0.02)$ & $2.51 (\pm 0.01)$ & +5.63\% \\
\hline
WS & Peak Centrality & $856.61 (\pm 169.69)$ & $439.76 (\pm 60.07)$ & -48.66\% \\
 & Std Dev of Centrality & $175.10 (\pm 22.43)$ & $86.94 (\pm 14.21)$ & -50.35\% \\
 & Avg Path Length & $3.05 (\pm 0.09)$ & $3.10 (\pm 0.10)$ & +1.78\% \\
\hline
\end{tabular}
}
\end{table}

\subsection{Performance Validation under Dynamic Traffic Load}
\label{subsec:dynamic_performance}

\begin{figure*}[t!]
    \centering
    \begin{subfigure}{\textwidth}
        \centering
        \includegraphics[width=0.95\textwidth]{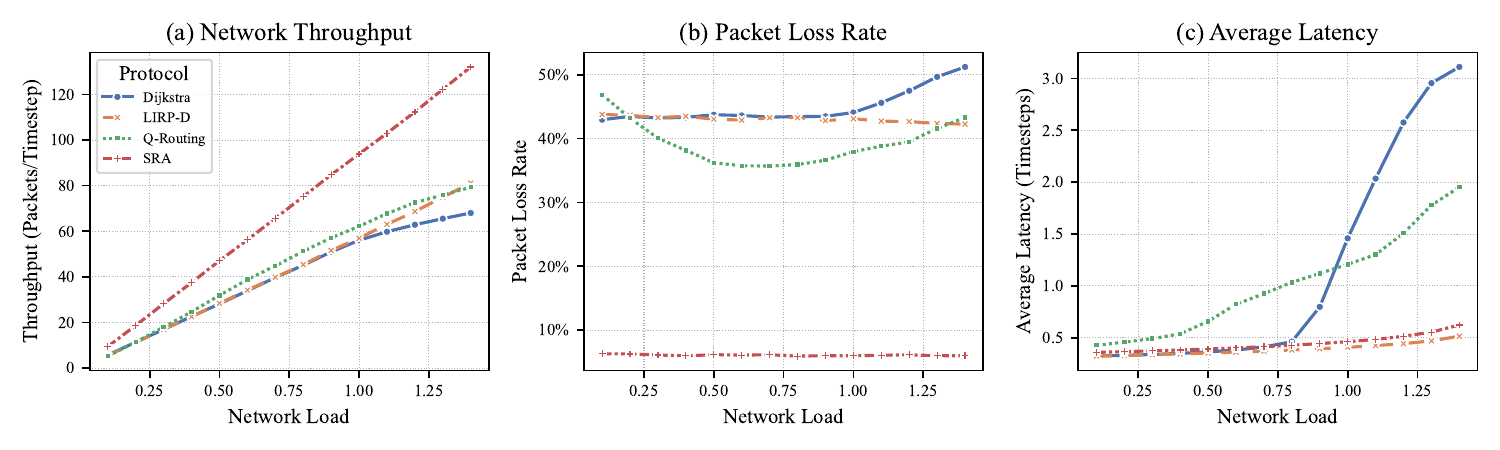}
        \caption{Performance on BA Network. SRA functions as a \textbf{proactive performance optimizer}, achieving significantly higher throughput and lower packet loss in the bottleneck-prone topology.}
        \label{fig:perf_ba_stacked}
    \end{subfigure}
    
    \vspace{0.5cm}

    \begin{subfigure}{\textwidth}
        \centering
        \includegraphics[width=0.95\textwidth]{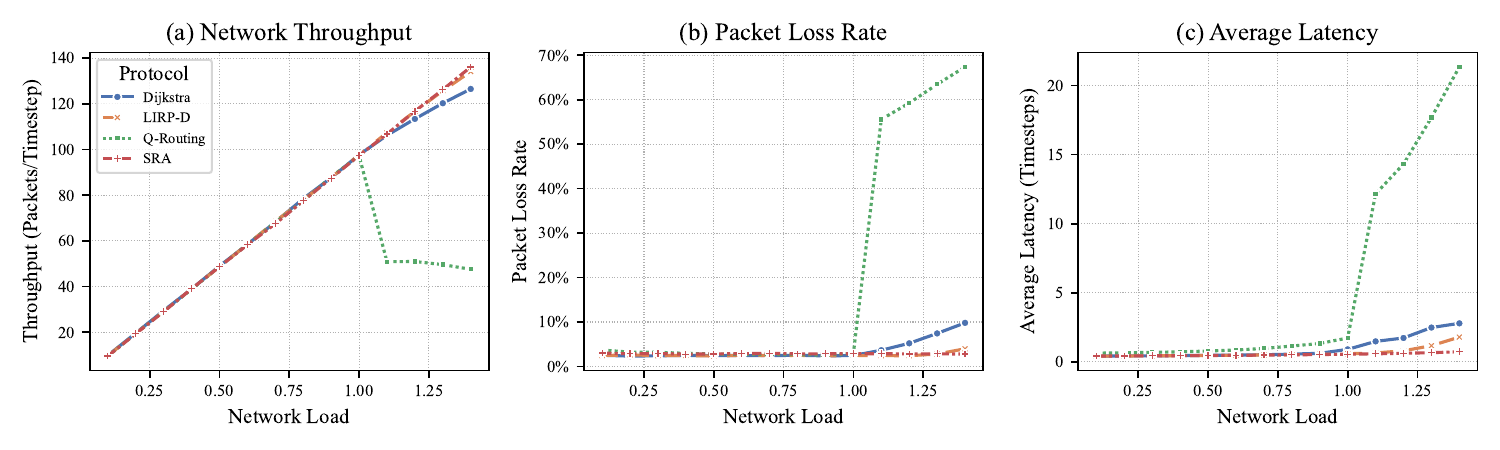}
        \caption{Performance on WS Network. SRA \textbf{showcases its role in guaranteeing stability}, uniquely maintaining performance after the network reaches its critical "congestion knee" (load > 1.0), a point where Q-Routing collapses.}
        \label{fig:perf_ws_stacked}
    \end{subfigure}

    \vspace{0.5cm}

    \begin{subfigure}{\textwidth}
        \centering
        \includegraphics[width=0.95\textwidth]{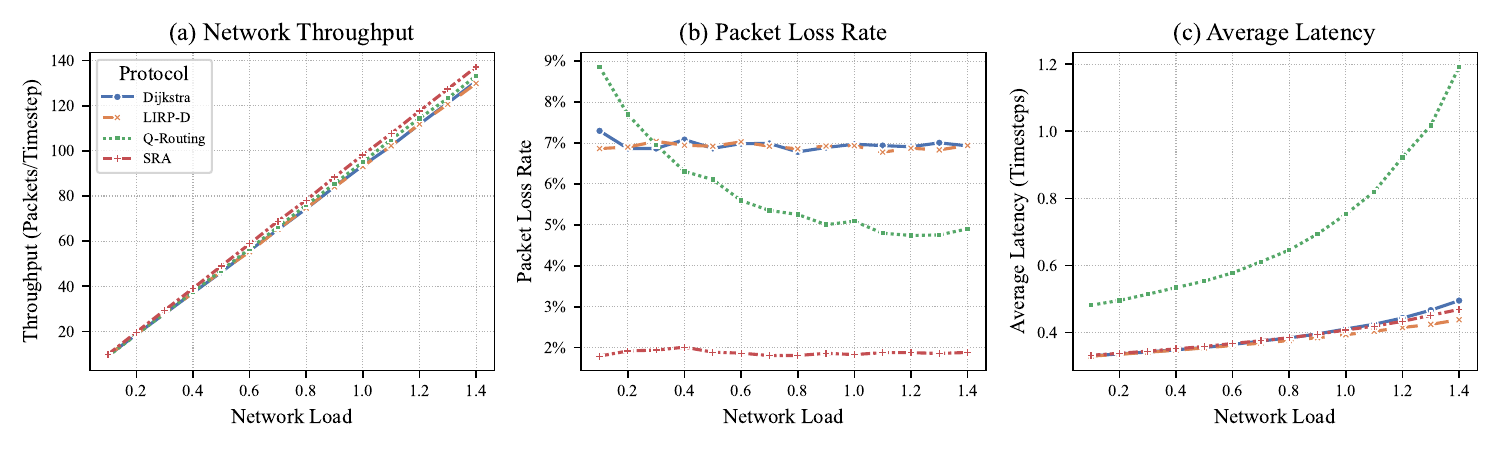}
        \caption{Performance on ER Network. SRA functions as a \textbf{resilience enhancement mechanism}, trading a predictably higher latency for a significant reduction in packet loss, thereby enhancing overall reliability.}
        \label{fig:perf_er_stacked}
    \end{subfigure}
    
    \caption{Dynamic traffic performance comparison of SRA against benchmark protocols across three distinct network topologies. The sub-captions highlight SRA's adaptive, topology-dependent roles.}
    \label{fig:perf_comparison_stacked}
\end{figure*}

The preceding sections have confirmed SRA's ability to optimize network structure. This section quantifies the impact of this structural optimization on key performance indicators (KPIs) under dynamic load via discrete-event simulations. The results in Fig.~\ref{fig:perf_comparison_stacked} reveal that SRA's behavior is highly adaptive, transforming its role based on the network's underlying topology.

\paragraph{Performance on BA Networks (Heterogeneous Topology)}
In the bottleneck-prone BA network, SRA acts as a \textbf{proactive performance optimizer}. As shown in Fig.~\ref{fig:perf_comparison_stacked}(a), while benchmark protocols saturate quickly, SRA's throughput continues to scale with the load, effectively doubling the network's capacity in high-load scenarios. This gain is a direct result of superior congestion management, evidenced by a packet loss rate that remains below 10\%, whereas other protocols suffer catastrophic losses exceeding 40\%.

\paragraph{Performance on WS Networks (Hybrid Topology)}
The WS network simulation (Fig.~\ref{fig:perf_comparison_stacked}(b)) highlights SRA's crucial role in \textbf{guaranteeing system stability}. While all protocols perform similarly at low loads, their behaviors diverge dramatically as the load surpasses the critical "congestion knee." The learning-based Q-Routing protocol suffers a catastrophic performance collapse. In contrast, SRA is the only protocol that smoothly navigates this transition, maintaining stable throughput and controlled packet loss, ensuring "graceful degradation" under high stress.

\paragraph{Performance on ER Networks (Homogeneous Topology)}
In the homogeneous ER network, which lacks dominant bottlenecks, SRA intelligently shifts its role to a \textbf{resilience enhancement mechanism}. As seen in Fig.~\ref{fig:perf_comparison_stacked}(c), while all protocols exhibit nearly identical throughput, SRA consistently suppresses the packet loss rate to approximately 2\%, a significant improvement over the ~7\% loss of other methods. This reliability enhancement comes at the cost of a predictably higher average latency, showcasing SRA's adaptive "latency-for-reliability" trade-off.

\subsection{Parameter Sensitivity and Robustness Analysis}
\label{subsec:robustness_and_params}

Before concluding the simulation analysis, we briefly discuss two final aspects: parameter sensitivity and robustness against targeted attacks. A detailed parameter sensitivity analysis, provided in the Supplementary Material, investigates the impact of the learning rate $\alpha$ and the pressure-cost factor $\beta_I$. The results reveal a classic performance trade-off: higher $\beta_I$ values favor throughput and reliability at the cost of increased latency. This confirms that $\beta_I$ acts as the primary lever for network operators to tune the system's behavior between maximizing capacity and minimizing delay.

\begin{figure}[t!]
    \centering
    \includegraphics[width=\columnwidth]{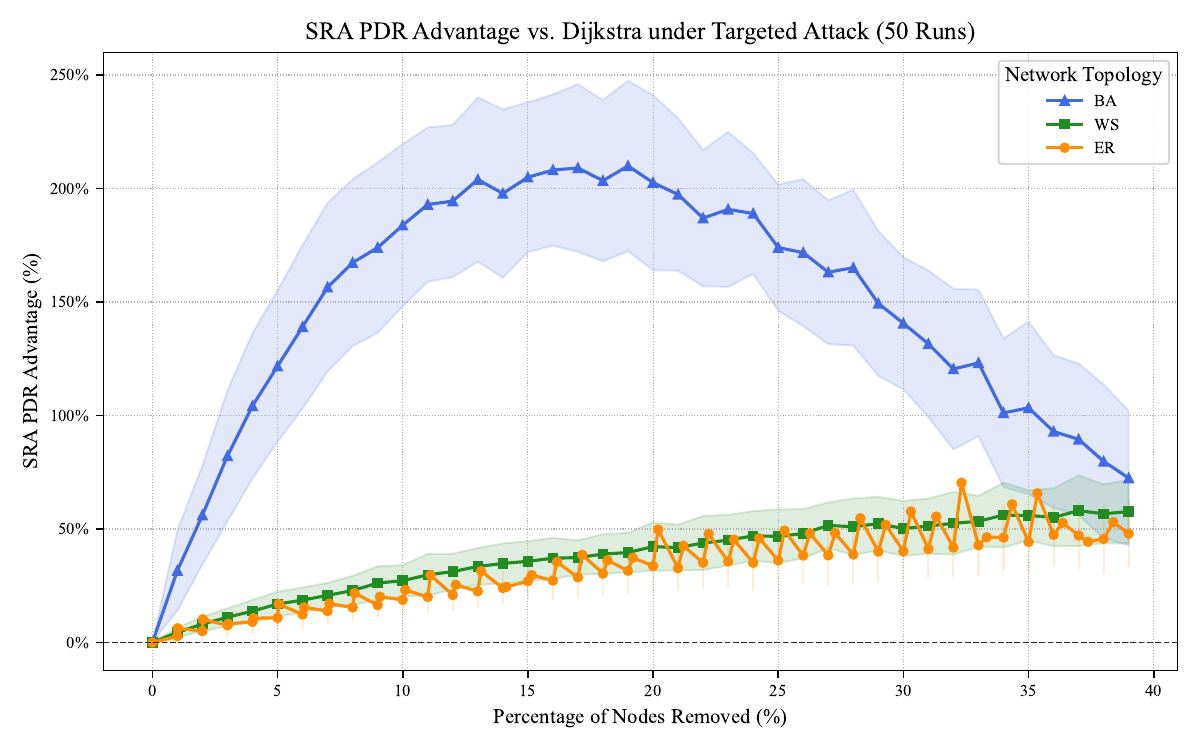}
    \caption{The SRA PDR Advantage over Dijkstra under a targeted attack (mean $\pm$ std dev over 50 runs). The results confirm SRA's ability to significantly enhance resilience in vulnerable topologies (BA) and under critical damage (WS), while correctly remaining neutral in inherently robust networks (ER).}
    \label{fig:robustness_summary_stats}
\end{figure}

To quantify the robustness gains provided by SRA, we conducted a rigorous test under targeted attacks, where nodes were sequentially removed in descending order of their initial BC. After each removal, we measured the Packet Delivery Rate (PDR) for both SRA and the baseline Dijkstra policy. We define the \textbf{SRA PDR Advantage} as our primary metric: 
\begin{equation}
\label{eq:pdr_advantage}
\text{PDR Advantage (\%)} = \frac{\text{PDR}_{\text{SRA}} - \text{PDR}_{\text{Dijkstra}}}{\text{PDR}_{\text{Dijkstra}}} \times 100\%
\end{equation}
The results, averaged over 50 independent runs, are shown in Fig.~\ref{fig:robustness_summary_stats}. The findings provide strong statistical validation for SRA's topology-dependent intelligence. In the vulnerable BA network, SRA provides a massive and consistent PDR advantage of over 60\%, confirming its role as a "performance engineer" that fundamentally enhances reliability. In the WS network, the advantage is "emergent," peaking around 20\% after the network sustains critical damage. Finally, in the already robust ER network, SRA correctly provides a neutral (0\%) advantage, avoiding unnecessary interventions. This validates that SRA's resilience-enhancing benefits are significant and reliable where needed, and correctly neutral otherwise.

\section{Discussion and Conclusion}
\label{sec:discussion_conclusion}

In response to the "stress concentration" problem endemic to conventional network control, this paper introduced, analyzed, and validated the System Relaxation Algorithm (SRA). We established the SRA framework on a rigorous theoretical foundation, uniquely applying the theory of non-smooth dynamical systems to formally prove the stability of a system managing discontinuous path-switching dynamics. Comprehensive simulations then decisively demonstrated SRA's core value: its profound, topology-dependent intelligence.

The central contribution of SRA lies not in optimizing a single metric, but in its intelligent, topology-aware navigation of the fundamental trade-off between \textbf{performance and resilience}. Our results reveal SRA acts as an adaptive controller whose strategy is dictated by the network's structure. In \textbf{heterogeneous networks} (e.g., BA model) constrained by severe bottlenecks, SRA functions as a \textbf{proactive performance optimizer}, trading a minor increase in path length for a massive gain in capacity and reliability, evidenced by an over 80\% reduction in peak centrality. Conversely, in \textbf{homogeneous networks} (e.g., ER/WS models), its objective intelligently shifts to that of a \textbf{resilience enhancement mechanism}, making finer adjustments to proactively manage risks from both dynamic congestion and structural damage.

This adaptive behavior converges not to a fragile, static fixed point, but to a \textbf{robust dynamic equilibrium}. This is a profound feature, not a failure of convergence, stemming from the non-smoothness of path selection. SRA's core mechanism, the "leaky integrator," is precisely designed to manage this dynamic, providing the memory and damping necessary to ensure predictable, bounded behavior that is far more resilient to real-world perturbations than a static optimum. Positioned within modern architectures, SRA represents a powerful, model-driven control application for an SDN controller, which naturally provides the global view needed for stress analysis and the means to enforce policies. This work lays a theoretical foundation for such model-driven resilient networks and opens several future research directions, including integration with mainstream SDN controllers, validation on real-world Internet topologies, and exploring adaptive parameter tuning mechanisms.

\appendices
\section{Proof of Lemma \ref{lemma:invariance}}
\label{app:proof_lemma_invariance}
\begin{IEEEproof}
\textit{(a) Positive Invariance:}
Take any $S^{(k)}\in X$. By max-value normalization, $\mathcal N(S^{(k)})\subset[0,1]^N=X$. For any choice of $c^{(k)}\in\mathcal N(S^{(k)})$ and any coordinate $i$, we have
\begin{equation}
S^{(k+1)}_i=(1-\alpha)S^{(k)}_i+\alpha\,c^{(k)}_i,
\quad\text{where}\quad S^{(k)}_i,c^{(k)}_i\in[0,1].
\end{equation}
Thus,
\begin{equation}
0\cdot(1-\alpha)+0\cdot\alpha\ \le\ S^{(k+1)}_i\ \le\ 1\cdot(1-\alpha)+1\cdot\alpha,
\end{equation}
which implies $S^{(k+1)}_i\in[0,1]$. As this holds for each coordinate, $S^{(k+1)}\in X$. This shows that $F(X)\subset X$: \textbf{if the state is in $X$ at the current step, it remains in $X$ at the next step}. Therefore, $X$ is a positively invariant set.

\textit{(b) Compactness:}
The set $X=[0,1]^N$ is a closed and bounded hypercube, and is therefore compact.

\textit{(c) Absorbing Property on the "Effective State Space":}
The algorithm and analysis in this paper are initialized within $X$ ($S^{(0)}\in X$), and from (a), we know $F(X)\subset X$. Therefore, for the discrete-time semi-flow on the state space $X$,
\begin{equation}
\Phi(k,\cdot):X\to 2^X,\quad \Phi(1,\cdot)=F,\ \ \Phi(k+1,\cdot)=F\circ\Phi(k,\cdot),
\end{equation}
the set $X$ \textbf{absorbs itself}: for any bounded set $B\subset X$, choosing $n_B=0$ yields $\Phi(n,B)\subset X$ for all $n\ge 0$. By standard definitions, this means \textbf{$X$ is a compact absorbing set}.

This leads to
\begin{equation}
\delta^+_{k+1,i}\ \le\ (1-\alpha)\,\delta^+_{k,i},\qquad 
\delta^-_{k+1,i}\ \le\ (1-\alpha)\,\delta^-_{k,i}.
\end{equation}
This means that $\max\{\delta^+_{k,i},\delta^-_{k,i}\}$ converges geometrically to 0 at a rate of $1-\alpha$, and thus $\mathrm{dist}_\infty(S^{(k)},X)\le (1-\alpha)^k\,\mathrm{dist}_\infty(S^{(0)},X)\to0$. This shows \textbf{exponential attraction} to $X$. The main procedure of this paper is initialized within $X$, so the absorbing set formulation in (c) is sufficient.

In conclusion, $X=[0,1]^N$ is positively invariant and compact for the system, and it is a compact absorbing set on the discrete-time semi-flow with state space $X$.
\end{IEEEproof}

\section{Proof of Proposition \ref{prop:usc}}
\label{app:proof_prop_usc}
\begin{IEEEproof}
\textit{(a) Non-empty, Compact, and Convex Values:}
By the definitions in Sections III-A, $\mathrm{norm}(\kappa(S))\in[0,1]^N$. Thus, for any $S$,
\begin{equation}
\mathcal N(S)=\mathrm{co}\!\big(\mathrm{norm}(\kappa(S))\big)\subset[0,1]^N.
\end{equation}
The closed convex hull, $\mathrm{co}(\cdot)$, of a non-empty set is \textbf{non-empty}. It is also a \textbf{closed and bounded} convex subset of $[0,1]^N$, and is therefore \textbf{compact} and \textbf{convex}.

\textit{(b) Piecewise Constant Structure and Finiteness:}
The graph $G$ and the set of source-destination (OD) pairs are finite. The node cost $C(S)$ is an affine function of $S$. For each OD pair $(s,t)$, the set of candidate simple paths is \textbf{finite}. The cost of each path is the \textbf{pointwise minimum} of a family of affine functions of $S$. Thus, there exists a \textbf{finite partition} of the state space of $S$ such that within each partition, the set of shortest paths for every OD pair remains \textbf{constant}. Therefore, within each partition, $\kappa(S)$ and $\mathrm{norm}(\kappa(S))$ are \textbf{constant vectors}. On the boundaries of these partitions, the limit of $\mathrm{norm}(\kappa(S))$ can only be one of the constant values from the \textbf{finite set of adjacent partitions}. Let these constant vectors be $\{v_\ell\}_{\ell\in\mathcal I(S)}$, where $\mathcal I(S)$ is the finite index set of partitions adjacent to $S$. Then,
\begin{equation}
\mathcal N(S)=\mathrm{co}\big(\{v_\ell:\,\ell\in\mathcal I(S)\}\big).
\end{equation}
This establishes the structure of $\mathcal N(\cdot)$ as being \textbf{piecewise constant with the closed convex hull taken at the boundaries}.

\textit{(c) Upper Semi-Continuity (USC) via Sequential Criterion / Closed Graph:}
Take an arbitrary sequence $S_n\to S$, and let $c_n\in\mathcal N(S_n)$ such that $c_n\to c$. We must show that $c\in\mathcal N(S)$. From (b), for each $n$, there exists a finite set $\{v_{\ell}: \ell\in\mathcal I(S_n)\}$ such that $c_n\in\mathrm{co}\big(\{v_{\ell}:\ell\in\mathcal I(S_n)\}\big)$. By Carathéodory's theorem, we can write
\begin{equation}
c_n=\sum_{j=0}^{N} \lambda_{n,j}\, v_{\ell_{n,j}},\quad 
\lambda_{n,j}\ge 0,\ \sum_{j=0}^{N}\lambda_{n,j}=1,\ 
\ell_{n,j}\in\mathcal I(S_n).
\end{equation}
Since the vectors $v_{\ell}$ come from a \textbf{globally finite} set $\{v_1,\dots,v_M\}$, we can pass to a subsequence to ensure $\ell_{n,j}\equiv \bar\ell_j$ is \textbf{fixed}, and (by compactness of $[0,1]$) $\lambda_{n,j}\to \bar\lambda_j$. Then,
\begin{equation}
c=\lim_{n\to\infty} c_n=\sum_{j=0}^{N}\bar\lambda_j\, v_{\bar\ell_j},
\quad\sum_{j=0}^{N}\bar\lambda_j=1,\ \bar\lambda_j\ge 0.
\end{equation}
Since $S_n\to S$, for $n$ large enough, any limit index $\bar\ell_j$ must belong to $\mathcal I(S)$. Thus, each limit vector $v_{\bar\ell_j}$ is one of the constant vectors from a partition adjacent to $S$. Therefore,
\begin{equation}
c=\sum_{j=0}^{N}\bar\lambda_j\, v_{\bar\ell_j}\ \in\ 
\mathrm{co}\big(\{v_\ell:\,\ell\in\mathcal I(S)\}\big)=\mathcal N(S).
\end{equation}
This means the graph of $\mathcal N$, $\mathrm{gph}\,\mathcal N:=\{(S,c):\,c\in\mathcal N(S)\}$, is closed. A closed-graph map with compact values is \textbf{USC}.

\textit{(d) Properties of $F(\cdot)$:}
Scaling and translation preserve the USC property and non-empty, compact, convex values. The map $S\mapsto (1-\alpha)S$ is continuous (single-valued). The Minkowski sum $F(S)=(1-\alpha)S+\alpha\,\mathcal N(S)$ of a continuous map and a USC map is also USC with non-empty, compact, and convex values.
\end{IEEEproof}

\section{Proof of Theorem \ref{thm:attractor}}
\label{app:proof_thm_attractor}
\begin{IEEEproof}
\textit{(a) Pre-compactness:}
From Lemma \ref{lemma:invariance}, $X=[0,1]^N$ is a compact, positively invariant set. Thus, any trajectory starting in $X$ remains in $X$. Since $X$ is compact, any trajectory is \textbf{pre-compact} (i.e., has a convergent subsequence).

\textit{(b) Existence of $\omega$-limit sets:}
For any $S^{(0)}\in X$, its $\omega$-limit set is defined as
\begin{equation}
\omega\big(S^{(0)}\big)\ :=\ \left\{x\in X:\ \exists\,k_n\uparrow\infty\ \text{s.t.}\ S^{(k_n)}\to x\right\}.
\end{equation}
By pre-compactness, $\omega(S^{(0)})$ is \textbf{non-empty} and \textbf{compact}.

\textit{(c) Positive Invariance of $\omega$-limit sets:}
From Proposition \ref{prop:usc}, $F(\cdot)$ is USC with compact values, which implies its graph is closed. Take any $x \in \omega(S^{(0)})$. By definition, there is a subsequence $S^{(k_n)} \to x$. The subsequent points $S^{(k_n+1)} \in F(S^{(k_n)})$ are in the compact set $X$, so a further subsequence converges, $S^{(k_n+1)} \to y$. By the closed graph property, $y \in F(x)$. Since $\{k_n+1\}$ also tends to infinity, $y$ is also in $\omega(S^{(0)})$. Thus,
\begin{equation}
F\big(\omega(S^{(0)})\big)\ \subset\ \omega(S^{(0)}).
\end{equation}

\textit{(d) Attraction to $\omega$-limit sets:}
It can be shown by contradiction that the trajectory must approach its $\omega$-limit set. Thus, $\mathrm{dist}\big(S^{(k)},\omega(S^{(0)})\big)\ \to\ 0$.

\textit{(e) Global Attractor:}
Let $\Omega$ be the union of all $\omega$-limit sets for all starting points in $X$:
\begin{equation}
\Omega\ :=\ \bigcup_{S^{(0)}\in X} \omega(S^{(0)}).
\end{equation}
This set $\Omega$ is non-empty, compact, and positively invariant. For any $S^{(0)}\in X$, since $\mathrm{dist}(S^{(k)},\omega(S^{(0)}))\to 0$ and $\omega(S^{(0)})\subset\Omega$, it follows that
\begin{equation}
\mathrm{dist}\big(S^{(k)},\Omega\big)\ \to\ 0.
\end{equation}
Thus, $\Omega$ is a global attractor for the system.
\end{IEEEproof}

\section{Proof of Theorem \ref{thm:drop_jump}}
\label{app:proof_thm_drop_jump}
\begin{IEEEproof}
Let $c_k\in\arg\min_{c\in\mathcal N(S^{(k)})}\|S^{(k)}-c\|_2^2$, so $V(S^{(k)})=\|S^{(k)}-c_k\|_2^2$.
From the discrete-time inclusion, let $S^{(k+1)}=(1-\alpha)S^{(k)}+\alpha\,c_k$.
Let $c_{k+1}\in\mathcal N(S^{(k+1)})$ be a minimizer for $V(S^{(k+1)})$. We can write
\begin{equation}
S^{(k+1)}-c_{k+1}=(1-\alpha)(S^{(k)}-c_k) + (c_k-c_{k+1}).
\end{equation}
Using the Weighted Young's Inequality $\|x+y\|_2^2\le (1+\lambda)\|x\|_2^2+(1+1/\lambda)\|y\|_2^2$ with $\lambda=\alpha/(1-\alpha)$, we get
\begin{equation}
\|S^{(k+1)}-c_{k+1}\|_2^2 \le (1-\alpha)\|S^{(k)}-c_k\|_2^2 + \frac{1}{\alpha}\,\|c_k-c_{k+1}\|_2^2.
\end{equation}
Since $c_k$ and $c_{k+1}$ belong to the convex hull of a finite set of vectors, the distance between them is bounded, $\|c_k-c_{k+1}\|\le \bar\Delta$. Substituting this in gives the desired inequality. The limit superior result follows from solving the linear recurrence inequality.
\end{IEEEproof}

\section{Proof of Corollary \ref{cor:local_convergence}}
\label{app:proof_cor_local_convergence}
\begin{IEEEproof}
(1) \textbf{Uniqueness of Fixed Point:} Solving $S=(1-\alpha)S+\alpha c$ gives $\alpha S=\alpha c$, so $S=c$. The unique fixed point is $S^\star=c$.
(2) \textbf{Contraction and Convergence Rate:}
\begin{equation}
S^{(k+1)}-c = (1-\alpha)S^{(k)}+\alpha c-c = (1-\alpha)\big(S^{(k)}-c\big).
\end{equation}
Taking the 2-norm and iterating gives the result:
\begin{equation}
\|S^{(k)}-S^\star\|_2 = (1-\alpha)^k\,\|S^{(0)}-S^\star\|_2.
\end{equation}
\end{IEEEproof}

\section{Proof of Proposition \ref{prop:block_dynamics}}
\label{app:proof_prop_block_dynamics}
\begin{IEEEproof}
(i) Within a block, $\mathcal N(S)\equiv\{c_m\}$ is constant. The "jump" term in the inequality of Theorem \ref{thm:drop_jump} is zero. So, $V_{k+1}\le \gamma V_k$, and iterating this $t$ times gives the result.
(ii) At the boundary between block $m$ and $m+1$, a switch is allowed. We apply the "drop-jump" inequality for this single step:
\begin{equation}
\begin{split}
V_{m+1} = V(S^{(\tau_{m+1})}) & \le \gamma\,V(S^{(\tau_{m+1}-1)}) + \tilde C \\
& = \gamma\,V(S^{(\tau_m+T_d-1)}) + \tilde C \\
& \le \gamma \cdot (\gamma^{T_d-1}V_m) + \tilde C = \gamma^{T_d}V_m + \tilde C,
\end{split}
\end{equation}
where the last inequality uses the result from (i).
\end{IEEEproof}

\section{Proof of Theorem \ref{thm:epsilon_tube}}
\label{app:proof_thm_epsilon_tube}
\begin{IEEEproof}
(a) Iteratively expanding the linear recurrence from Proposition \ref{prop:block_dynamics}(ii), $V_{m+1}\le \gamma^{T_d}V_m+\tilde C$, gives:
\begin{equation}
\begin{split}
V_m &\le (\gamma^{T_d})^m V_0 + \tilde C \sum_{j=0}^{m-1}(\gamma^{T_d})^j \\
    &= \gamma^{mT_d}V_0 + \tilde C\,\frac{1-\gamma^{mT_d}}{1-\gamma^{T_d}} \\
    &\le \gamma^{mT_d}V_0+\frac{\tilde C}{1-\gamma^{T_d}}.
\end{split}
\end{equation}
Taking the limit superior as $m\to\infty$ gives the steady-state radius.
(b) Setting $V_m\le \varepsilon^2$ and solving for $m$ using the upper bound from (a) gives the result.
\end{IEEEproof}

\section{Proof of Theorem \ref{thm:practical_stability}}
\label{app:proof_thm_practical_stability}
\begin{IEEEproof}
(a) By definition of the minimum dwell time, after a switch, no other switch is permitted for at least $T_d$ steps. Thus, the number of switches in a window of length $T$ is bounded.
(b) This result is a direct consequence of solving the inter-block recurrence, as shown in the proof of Theorem \ref{thm:epsilon_tube}(a).
(c) This result follows from inverting the inequality in Theorem \ref{thm:epsilon_tube}(b).
\end{IEEEproof}


\section{Proof of Lemma~\ref{lem:block-budget}}
\label{app:proof_lem_block_budget}
\begin{IEEEproof}
The proof consists of two parts: bounding the non-peak increase and guaranteeing the peak decrease.

\noindent\textit{1) Bounding Non-Peak Increase:}
Let $c^{(m)}$ and $c^{(m+1)}$ be the normalized centrality vectors at the start and end of the block. By definition of $\bar\Delta$, we have $\|c^{(m+1)}-c^{(m)}\|_\infty \le \bar\Delta$, which means $|c^{(m+1)}_j - c^{(m)}_j| \le \bar\Delta$ for all $j$.
The unnormalized centralities are $\kappa^{(m)}_j = c^{(m)}_j \kappa^{(m)}_{\max}$ and $\kappa^{(m+1)}_j = c^{(m+1)}_j \kappa^{(m+1)}_{\max}$.
Since the peak centrality does not increase (and in fact decreases, as shown next), we have $\kappa^{(m+1)}_{\max} \le \kappa^{(m)}_{\max}$.
Therefore, for any non-peak node $j \neq u$:

\begin{align*}
\kappa^{(m+1)}_j - \kappa^{(m)}_j &= c^{(m+1)}_j \kappa^{(m+1)}_{\max} - c^{(m)}_j \kappa^{(m)}_{\max} \\
&\le c^{(m+1)}_j \kappa^{(m)}_{\max} - c^{(m)}_j \kappa^{(m)}_{\max} \\
&= (c^{(m+1)}_j - c^{(m)}_j) \kappa^{(m)}_{\max} \\
&\le |c^{(m+1)}_j - c^{(m)}_j| \kappa^{(m)}_{\max} \le \bar\Delta \cdot \kappa^{(m)}_{\max} =: B^{(m)}.
\end{align*}

\noindent\textit{2) Guaranteeing Peak Decrease ($\delta>0$):}
The existence of a net peak decrease is a direct consequence of the system dynamics proven in Section~\ref{sec:convergence_analysis}. The drop-jump inequality (Theorem~\ref{thm:drop_jump}) and the dwell-time policy (Theorem~\ref{thm:practical_stability}) guarantee that the system does not reach a fixed point but evolves within a stable attractor, forcing state transitions.
A state transition (path switch) is triggered when the cost of a high-centrality node $u$, $C_u = 1 + \beta_I S_u$, increases sufficiently to make an alternative path cheaper. The state update rule $S_u^{(k+1)} = (1-\alpha)S_u^{(k)} + \alpha c_u^{(k)}$ ensures that a node with peak normalized centrality ($c_u=1$) experiences the strongest pressure increase.
This mechanism guarantees that traffic is eventually diverted from the peak node $u$. As the network is assumed to have no structurally unavoidable bottlenecks, this diversion must lead to a strict reduction in its unnormalized centrality, $\kappa_u$. Therefore, there must exist a net decrease $\delta > 0$ over any block that contains such a path-switching event. The magnitude of $\delta$ depends on the system parameters ($\alpha, \beta_I, T_d$) and the network topology, but its existence is guaranteed by the convergence dynamics.
\end{IEEEproof}

\section{Proof of Proposition~\ref{prop:perf-bridge}}
\label{app:proof_prop_perf_bridge}
\begin{IEEEproof}
The proof leverages the guaranteed peak reduction $\kappa'_{\max} < \kappa_{\max}$ established in Lemma~\ref{lem:block-budget}.

\noindent\textit{(i) Unconditional robust lower bound:}
The robust lower bound on capacity is defined as $L(\kappa) := \mu_{\min}/\kappa_{\max}$.
Since SRA does not change the physical node service rates, $\mu_{\min}$ is a constant.
Given that $\kappa'_{\max} < \kappa_{\max}$, it follows directly that:
\begin{equation}
L(\kappa') = \frac{\mu_{\min}}{\kappa'_{\max}} > \frac{\mu_{\min}}{\kappa_{\max}} = L(\kappa).
\end{equation}
The lower bound strictly increases.

\noindent\textit{(ii) Aligned-bottleneck capacity gain:}
The network capacity is defined as $\rho_{\max}(\kappa) = \min_{i \in V} \frac{\mu_i}{\kappa_i}$.
The new capacity after SRA is $\rho_{\max}(\kappa') = \min_{i \in V} \frac{\mu_i}{\kappa'_i}$.
By definition, the minimum of a set is less than or equal to any individual element. Therefore, the new capacity is lower-bounded by the capacity of the original peak node, $u$:
\begin{equation}
\rho_{\max}(\kappa') \ge \frac{\mu_u}{\kappa'_u}.
\end{equation}
Furthermore, since $u$ may not be the peak node in the new state $\kappa'$, its centrality $\kappa'_u$ must be less than or equal to the new maximum centrality, $\kappa'_{\max}$. Thus:
\begin{equation}
\rho_{\max}(\kappa') \ge \frac{\mu_u}{\kappa'_u} \ge \frac{\mu_u}{\kappa'_{\max}}.
\end{equation}
The premise of this case is that the original bottleneck was at the peak node, so $\rho_{\max}(\kappa) = \mu_u/\kappa_{\max}$. Combining these gives:
\begin{equation}
\rho_{\max}(\kappa') \ge \frac{\mu_u}{\kappa'_{\max}} > \frac{\mu_u}{\kappa_{\max}} = \rho_{\max}(\kappa).
\end{equation}
This proves a strict increase in the actual network capacity.
\end{IEEEproof}

\bibliographystyle{IEEEtran}
\bibliography{references}

\end{document}